\documentstyle[aps,eqsecnum,12pt]{revtex}

\begin{document}

\title{Propagation of gravitational waves from slow motion sources \\ 
in a Coulomb type potential} 

\author{Hideki ASADA \footnote{Present address: 
Faculty of Science and Technology \\
Hirosaki University, 
Hirosaki 036, JAPAN} \\
Yukawa Institute for Theoretical Physics, 
Kyoto University, 
Kyoto 606-01, Japan \\
and \\
Toshifumi FUTAMASE \\ 
Astronomical Institute, Graduate School of 
Science, Tohoku University \\
Sendai 980-77, Japan}

\maketitle

\vspace{1cm}

\begin{abstract}
We consider the propagation of gravitational waves generated 
by slow motion sources in Coulomb type potential 
due to the mass of the source. 
Then, the formula for gravitational waveform including tail 
is obtained in a straightforward manner by using 
the spherical Coulomb function. 
We discuss its relation with the formula in the previous work. 
\end{abstract}

\begin{flushleft}
PACS Number(s): 04.25.Nx, 04.30.Nk 
\end{flushleft}

\def\pa{\partial}
\def\bI{\hbox{$\,I\!\!\!\!-$}}
\def\two{\hbox{$_{(2)}$}}
\def\three{\hbox{$_{(3)}$}}
\def\four{\hbox{$_{(4)}$}}
\def\five{\hbox{$_{(5)}$}}
\def\six{\hbox{$_{(6)}$}}
\def\seven{\hbox{$_{(7)}$}}
\def\eight{\hbox{$_{(8)}$}}
\newcommand{\lsim}{\raisebox{0.3mm}{\em $\, <$} \hspace{-3.3mm}
\raisebox{-1.8mm}{\em $\sim \,$}}

\newpage

\section{Introduction}

It becomes urgent to study the generation and the propagation 
of gravitational waves in detail, because of the increasing expectation 
of direct detection of gravitational waves by Kilometer-size 
interferometric gravitational wave detectors, such as LIGO\cite{abramovici} 
and VIRGO\cite{bradaschia} now under construction. 
Coalescing binary neutron stars are the most promising candidate of 
sources of gravitational waves for such detectors.  
Gravitational waves generated by them bring us informations not only on 
various physical parameters of neutron star\cite{thorne87,thorne94}, 
but also on the cosmological parameters\cite{schutz86,mark,finn,wst} 
if and only if we can make a detailed comparison 
between the observed signal with theoretical prediction during 
the epoch of the so-called inspiraling phase where 
the orbital separation is much larger than the radius of component stars 
\cite{cab}.
The problem is that, in order to make any successful comparison between 
theory and observation, we need to know the detailed waveform generated by  
the motion up to 4PN order \cite{tn} which is of order 
$\epsilon^8$ higher than the Newtonian order, where 
$\epsilon=\mbox{orbital velocity}/\mbox{light speed}$. 

Blanchet and Damour have developed a systematic scheme 
to calculate the waveform at higher orders, 
where the post-Minkowskian approximation is used to construct 
the external field and the post-Newtonian approximation is used 
to construct the field near the material source 
\cite{bd86,bd88,bd92}. 
In the post-Minkowskian approximation, the background geometry is 
the Minkowski spacetime where linearized gravitational waves propagate 
\cite{thorne80,bd86,bd88}. 
The corrections to propagation of gravitational waves 
can be taken into account by performing the post-Minkowskian 
approximation up to higher orders. 
In fact, Blanchet and Damour obtained the tail term 
of gravitational waves as the integral over the past history 
of the source \cite{bd88,bd92}. 
They introduced a complex parameter $B$ \cite{bd86}, and used 
the analytic continuation as a useful mathematical device 
to evaluate the so-called $\log$ term in the tail
contribution \cite{bd88,bd92}. 
The method is powerful, but it is not easy to see the origin of 
the tail term. 
Will and Wiseman \cite{ww96} have also obtained the tail term 
associated with the mass quadrupole moment, by generalizing 
the Epstein-Wagoner formalism \cite{ew}. 
In the following, we shall study the waveform formula 
from different point of view because of its importance. 
Namely we shall use the Green's function for the wave operator 
in the Coulomb type potential generated by the mass of the source. 
This method is straightforward and enables us to see explicitly 
how the tail term originates 
from the difference 
between the flat light cone and the true one, which is due to 
the mass of the source $GM/c^2$ as the lowest order correction. 
Sch\"afer\cite{schafer90} and Nakamura\cite{nakamura95} 
have obtained the quadrupole energy loss formula 
including the contribution from the tail term by studying 
the wave propagation in the Coulomb type potential.

This paper is organized as follows. 
In section 2, we derive the formula for the tail of gravitational waves 
by using the Green's function in the Coulomb type potential. 
In section 3, we discuss its relation with previous works. 
We use the units of $c=G=1$.

\section{Gravitational waves in Coulomb type potential} 

We wish to clarify that the main part of 
the tail term is due to the propagation of gravitational waves 
on the light cone which deviates slightly from the flat light cone 
owing to the mass of the source. 
For this purpose, we shall work in the harmonic coordinate since 
the deviation may be easily seen in the reduced Einstein's equation 
in this coordinate. 
\begin{equation}
 ( \eta^{\alpha\beta} - {\bar h}^{\alpha\beta} )
{\bar h}^{\mu\nu}_{\;\;\; ,\alpha\beta} 
= - 16 \pi \Theta^{\mu\nu} 
+ {\bar h}^{\mu\alpha}_{\;\;\; ,\beta}
{\bar h}^{\nu\beta}_{\;\;\; ,\alpha} , 
\label{waveeq0}
\end{equation}
where $\eta^{\mu\nu}$ is the Minkowskian metric, 
$\bar h^{\mu\nu}=\eta^{\mu\nu}-\sqrt{-g} g^{\mu\nu}$
and 
\begin{equation}
\Theta^{\mu\nu} = (-g)(T^{\mu\nu}+t^{\mu\nu}_{LL}) . 
\end{equation}
Here, $t^{\mu\nu}_{LL}$ is the Landau-Lifshitz pseudotensor \cite{ll}. 
Since we look for  $1/r$ part of the solution, the spatial derivative 
is not relevant in the differential operator. 
Therefore, it is more convenient for our purpose to transform 
Eq.($\ref{waveeq0}$) into the following form 
\begin{equation}
( \Box - {\bar h}^{00}\partial_0 \partial_0 )
{\bar h}^{\mu\nu}= - 16 \pi {\cal S}^{\mu\nu} , 
\end{equation}
where $\Box=\eta^{\mu\nu} \partial_{\mu}\partial_{\nu}$ and 
we defined ${\cal S}^{\mu\nu}$ as 
\begin{equation}
{\cal S}^{\mu\nu} = \Theta^{\mu\nu} 
-{1\over {16\pi}} ( {\bar h}^{\mu\alpha}_{\;\;\; ,\beta} 
{\bar h}^{\nu\beta}_{\;\;\; ,\alpha}
+ 2{\bar h}^{0i}{\bar h}^{\mu\nu}_{\;\;\; ,0i} 
+ {\bar h}^{ij}{\bar h}^{\mu\nu}_{\;\;\; ,ij} ) . 
\label{waveeq1}
\end{equation}
Substituting the lowest order expression for ${\bar h}^{00}$, 
we obtain our basic equation 
\begin{equation}
\Box_M {\bar h}^{\mu\nu} = \tilde\tau^{\mu\nu} ,  
\label{tildewaveeq}
\end{equation}
where the effective source $\tilde \tau^{\mu\nu}$ is defined as 
\begin{equation}
{\tilde \tau^{\mu\nu}} 
=  {\cal S}^{\mu\nu} 
- {1\over {16\pi}} \left({\bar h}^{00} 
- {{4M}\over r} \right){\bar h}^{\mu\nu}_{\;\;\; ,00} , 
\end{equation}
Here, $M$ is the total mass of the source and $\Box_M$ is defined as 
\begin{equation}
\Box_M=\Bigl[ -\Bigl( 1+{4M \over r} \Bigr) {\partial^2 \over {\partial t^2}}
+\Delta \Bigr], 
\label{box2}
\end{equation}
where $\Delta$ is the Laplacian in the flat space. 
In case of the source with strong internal gravity like 
binary neutron stars, $M$ may be regarded as the ADM mass \cite{futamase87}, 
since $1/r$ part of $\bar h^{00}$ in the harmonic coordinate 
becomes the ADM mass. 
In this way, our derivation may be applied to such compact sources. 

The solution for $(\ref{tildewaveeq})$ may be written down by using the 
retarded Green's function in the following form 
\begin{equation}
{\bar h}^{\mu\nu}(x) = \int d^4 x' G_M^{(+)}(x, x') 
{\tilde  \tau}^{\mu\nu}(x') . 
\label{formalsol}
\end{equation}
The retarded Green's function is defined as satisfying the equation; 
\begin{equation}
\Box_M G_M^{(+)}(x,y)= \delta^4(x-y) , \label{curvedG}
\end{equation}
with an appropriate boundary condition. 

The Green's function satisfying Eq.($\ref{curvedG}$) can be constructed 
by using the homogeneous solutions for the equation 
\begin{equation}
\Box_M\Psi=0 . 
\label{homoeq}
\end{equation}
The homogeneous solution for Eq.($\ref{homoeq}$) takes a form of 
\begin{equation}
e^{-i\omega t}f_l(\rho)Y_{l m}(\theta,\phi) , 
\label{homsol}
\end{equation}
where we defined 
\begin{equation}
\rho=\omega r . 
\end{equation}
Then the radial function $\tilde f_l(\rho)\equiv \rho f_l(\rho)$ 
satisfies  
\begin{equation}
\Bigl( {d^2 \over d\rho^2}+1+{4M\omega \over \rho}-{l(l+1) \over \rho^2} 
\Bigr) \tilde f_l(\rho)=0 , 
\end{equation}
so that Eq.($\ref{homsol}$) is a solution for Eq.($\ref{homoeq}$). 
Thus we can obtain homogeneous solutions for Eq.($\ref{homoeq}$) 
by choosing $\tilde f_l(\rho)$ as one of spherical Coulomb functions; 
$u^{(\pm)}_l(\rho;\gamma)$ and $F_l(\rho;\gamma)$ 
with $\gamma=-2M\omega$. 
Here, we adopted the following definition of the spherical Coulomb 
function \cite{messiah} as 
\begin{eqnarray}
F_l(\rho;\gamma)&=&c_le^{i\rho}\rho^{l+1}F(l+1+i\gamma|2l+2|-2i\rho) , 
\nonumber\\ 
u^{(\pm)}_l(\rho;\gamma)&=&\pm 2ie^{\mp i\sigma_l} c_l 
e^{\pm i\rho}\rho^{l+1} 
W_1(l+1\pm i\gamma|2l+2|\mp 2i\rho) , 
\end{eqnarray}
where $c_l$ and $\sigma_l$ are defined as 
\begin{eqnarray}
c_l&=&2^l e^{-\pi\gamma /2}{|\Gamma(l+1+i\gamma)| \over (2l+1)!} , 
\nonumber\\
\sigma_l&=&\mbox{arg}\Gamma(l+1+i\gamma) . 
\label{cl}
\end{eqnarray}
Here, $F$ and $W_1$ are the confluent hypergeometric function and 
the Whittaker's function respectively. 
These spherical Coulomb functions have asymptotic behavior as 
\begin{eqnarray}
u^{(\pm)}_l&\sim&\exp\Bigl[ \pm i\Bigl( \rho-\gamma\ln{2\rho}
-{1 \over 2}l\pi \Bigr) \Bigr] \quad\mbox{for}\quad r \to \infty , 
\label{coulomb1}
\end{eqnarray}
and 
\begin{eqnarray}
F_l&\sim&c_l \rho^{l+1} \quad\mbox{for}\quad r \to 0 . 
\label{coulomb2}
\end{eqnarray}
Thus we obtain the retarded Green's function in the following form. 
\begin{eqnarray}
G^{(+)}_M(x,x')&=&\sum_{l m} e^{i \sigma_l} 
\int d\omega sgn(\omega) \Bigl( 
\Psi^{+ \,\omega\,l\,m}(x) \Psi^{S\,\omega\,l\,m\,\ast}(x') 
\theta(r-r') \nonumber\\
&&~~~~~~~~~~~~~~~~~~~~~~~~~~
+\Psi^{S\,\omega\,l\,m}(x) \Psi^{+ \,\omega\,l\,m\,\ast}(x') 
\theta(r'-r) \Bigr) , 
\label{green}
\end{eqnarray}
where we defined $\Psi^{+ \,\omega\,l\,m}(x)$ and
$\Psi^{S\,\omega\,l\,m}(x)$ as 
\begin{eqnarray}
\Psi^{+\omega\,l\,m}(x)&=&\sqrt{{|\omega| \over 2\pi}}
e^{-i\omega t} \rho^{-1} u^{(+)}_{l}(\rho;\gamma)
Y_{l\,m} , \nonumber\\
\Psi^{S\,\omega\,l\,m}(x)&=&\sqrt{{|\omega| \over 2\pi}} 
e^{-i\omega t} \rho^{-1} F_{l}(\rho;\gamma) 
Y_{l\,m} . 
\end{eqnarray}

For slow motion sources, we evaluate the asymptotic form
of the Green's function up to $O(M\omega)$ as 
\begin{eqnarray}
G^{(+)}_M(x,x')&=&{1 \over r}\sum_{lm}{(-i)^l \over 2\pi (2l+1)!!} 
\int d\omega \Bigl\{ 1+\pi M\omega+2iM\omega 
\Bigl( \ln{2M\omega}-\sum^l_{s=1} {1 \over s} +C \Bigr) 
\nonumber\\
&&+O(M^2\omega^2) \Bigr\}  e^{-i \omega(t-r-t')} (\omega r')^l 
Y_{lm}(\Omega) Y^{*}_{lm}(\Omega') +O(r^{-2}) \nonumber\\
&=&{1 \over r} \sum_{lm}{(-i)^l \over 2\pi (2l+1)!!}
\int d\omega \Bigl[ 1+2M\omega \Bigl\{ i \Bigl( -\sum^l_{s=1}{1 \over s} 
+\ln{2M} \Bigr)+{\pi \over 2} sgn(\omega)
\nonumber\\ 
&&+i (\ln{|\omega|}+C) \Bigr\} +O(M^2\omega^2) \Bigr] 
e^{-i \omega(t-r-t')} (\omega r')^l 
Y_{lm}(\Omega) Y^{*}_{lm}(\Omega') 
\nonumber\\ 
&&+O(r^{-2}) , 
\label{green3}
\end{eqnarray}
where $C$ is Euler's number. In deriving the above expression, we have used 
Eqs.($\ref{coulomb1}$) and ($\ref{coulomb2}$) and 
the following expansion for $c_l$ and $\sigma_l$ in $M\omega$ 
\begin{eqnarray}
c_l&=&{1+\pi M\omega+O(M^2\omega^2) \over (2l+1)!!} , \nonumber\\
\sigma_l&=&2M\omega \Bigl( C-\sum^{l}_{s=1}{1 \over s} \Bigr) 
+O(M^2\omega^2) . 
\end{eqnarray}

Now we apply the formula \cite{gr,bs} 
\begin{equation}
\omega\int_0^1 dv e^{i\omega v} \ln{v}+i\int_1^{\infty}
{dv \over v} e^{i\omega v}
=-{\pi \over 2}sgn(\omega)-i(\ln{|\omega|}+C), 
\label{formula} 
\end{equation}
to Eq.($\ref{green3}$), then we obtain
\begin{eqnarray}
G^{(+)}_M(x,x')&=&{1 \over r} \sum_{lm}{(-i)^l \over 2\pi (2l+1)!!}
\int d\omega 
\Bigl\{ 1-2M \Bigl( -\sum^l_{s=1}{1 \over s} +\ln{2M} \Bigr) {d \over dt} 
+2M \Bigl(\int^1_0 dv e^{i\omega v} \ln{v} \Bigr) {d^2 \over dt^2} 
\nonumber\\
&&\hspace*{1.5cm}+2M \Bigl(\int^{\infty}_1 {dv \over v} e^{i\omega v} 
\Bigr) {d \over dt} 
+O(M^2\omega^2) \Bigr\} \nonumber\\
&&\times e^{-i \omega(t-r-t')} (\omega r')^l 
Y_{lm}(\Omega) Y^{*}_{lm}(\Omega') +O(r^{-2}) . 
\label{green4}
\end{eqnarray}

Here, we assume the no incoming radiation condition 
on the initial hypersurface so that we may take  
\begin{equation}
\lim_{v \to \infty} e^{-i \omega (t-r-v-t')} \ln{v} \to 0 . 
\end{equation}
Thus we can make the following replacement  
\begin{equation}
\int^{\infty}_1 {dv \over v} e^{i \omega v} \to 
-i \omega \int^{\infty}_1 dv e^{i \omega v} \ln{v}  
=\Bigl( \int^{\infty}_1 dv e^{i \omega v} \ln{v} \Bigr) {d\over dt} . 
\label{replace}
\end{equation}
Inserting Eq.($\ref{replace}$) into Eq.($\ref{green4}$), we finally 
obtain the desired expression  for the retarded Green's function
\begin{eqnarray}
G^{(+)}_M(x,x')&=&{1 \over r} \sum_{lm}{(-i)^l \over 2\pi (2l+1)!!} 
\int d\omega 
\Bigl[ 1+2M \Bigl( \sum^l_{s=1}{1 \over s} -\ln{2M} \Bigr) {d \over dt}
\nonumber\\
&&\hspace*{2cm}+2M \Bigl(\int^{\infty}_0 dv e^{i\omega v} \ln{v}
\Bigr) {d^2 \over dt^2} +O(M^2 \omega^2) \Bigr] 
\nonumber\\
&&\times e^{-i \omega(t-r-t')} (\omega r')^l 
Y_{lm}(\Omega) Y^{*}_{lm}(\Omega') +O(r^{-2}) 
\nonumber\\ 
&=&{1 \over r} \;\mbox{part of} \nonumber\\ 
&&\Bigl[ G_{0}(x,x')+2M {d^2 \over dt^2} \sum_{lm} 
\int dv \Bigl\{ \ln \Bigl( {v \over 2M} \Bigr) 
+\sum^l_{s=1} {1 \over s} \Bigr\} 
G_{0}^{l\,m}(t-v,{\bf x},x') 
\nonumber\\
&&+O(M^2) \Bigr] +O(r^{-2}) , 
\label{green5}
\end{eqnarray}
where we defined the spherical harmonic expansion coefficient of the 
flat Green's function as 
\begin{equation}
G_{0}^{l\,m}(x,x')=Y_{l\,m}(\Omega') \int d\Omega' G_{0}(x,x') 
Y_{l\,m}(\Omega') . 
\end{equation}
As a result, we obtain the waveform generated by the effective source 
as 
\begin{eqnarray}
h_{ij}^{TT}&=&{4 \over r} P_{ijpq} 
\sum^{\infty}_{l=2} {1 \over l!} 
\Bigl[ n_{L-2} \Bigl\{ \tilde M_{pqL-2}(t-r)
+2M {d^2 \over dt^2} 
\int dv \Bigl\{ \ln \Bigl({v \over 2M} \Bigr) 
+\sum^{l-2}_{s=1} {1 \over s} \Bigr\} 
\nonumber\\
&&\hspace*{2.5cm}\times \tilde M_{pqL-2}(t-r-v) \Bigr\} 
-{2l \over l+1} n_{aL-2} 
\Bigl\{ \epsilon_{ab(p} \tilde S_{q)bL-2}(t-r)
\nonumber\\
&&\hspace*{2.5cm}+2M {d^2 \over dt^2} \int dv 
\Bigl\{ \ln \Bigl({v \over 2M}\Bigr)
+\sum^{l-1}_{s=1} {1 \over s} \Bigr\}  
\epsilon_{ab(p} \tilde S_{q)bL-2}(t-r-v) \Bigr\} 
\nonumber\\
&&\hspace*{2.5cm}+O(M^2) \Bigr] 
+O(r^{-2}) , 
\label{waveform}
\end{eqnarray}
where $P_{ijpq}$ is the transverse and traceless projection tensor 
and parentheses denote symmetrization, and $n_L$ is the tensor product of 
$L$ radial unit vectors. 
Although this waveform corresponds to (C1) by Blanchet \cite{blan95}, 
$\tilde M_{pq L-2}$ and $\tilde S_{pq L-2}$ are 
the mass and current multipole moments generated 
by the full nonlinear effective source $\tilde\tau^{\mu\nu}$. 
They take the same forms as in \cite{thorne80}, but here 
$\tilde\tau^{\mu\nu}$ is used in place of $\tau^{\mu\nu}$ in 
\cite{thorne80}, where 
\begin{equation}
\tau^{\mu\nu}=\Theta^{\mu\nu}-{1 \over 16\pi} 
\Bigl( \bar h^{\mu\alpha}_{\;\;\; ,\beta} 
\bar h^{\nu\beta}_{\;\;\; ,\alpha}
+\bar h^{\alpha\beta} \bar h^{\mu\nu}_{\;\;\; ,\alpha\beta} \Bigr) . 
\end{equation} 
It is worthwhile to point out that $\ln{2M}$ in Eq.($\ref{waveform}$) 
can be removed by using the freedom to time translation.

\section{Comparison with the previous work} 

\subsection{waveform in the post-Minkowskian approximation} 

By using the post-Minkowskian approximation, Blanchet obtained 
the general formula for gravitational waves including tail 
\cite{blan95}, 
\begin{eqnarray}
h_{ij}^{TT}&=&{4 \over r} P_{ijpq} 
\sum^{\infty}_{l=2} {1 \over l!} 
\Bigl[ n_{L-2} U_{pqL-2}(t-r)
-{2l \over l+1} n_{aL-2} 
\epsilon_{ab(p}V_{q)bL-2}(t-r) \Bigr] 
+O(r^{-2}) ,
\label{tailwave}
\end{eqnarray}
where the radiative mass moment and radiative current moment 
are defined as 
\begin{eqnarray}
U^{(l)}_L(u)&=&M^{(l)}_L(u)+2GM\int^{\infty}_0 dv M^{(l+2)}_L(u-v)
\Big\{ \ln \Bigl({v \over P}\Bigr)+\kappa_l \Bigr\}  
+O(G^2M^2) , 
\label{blanmoment}
\end{eqnarray}
\begin{eqnarray}
V^{(l)}_L(u)&=&S^{(l)}_L(u)+2GM\int^{\infty}_0 dv S^{(l+2)}_L(u-v)
\Big\{ \ln \Bigl({v \over P}\Bigr)+\kappa_l^{\prime} \Bigr\} 
+O(G^2M^2) . 
\label{blanmoment2}
\end{eqnarray}
Here the moments $M_L$ and $S_L$ do not contain the nonlinear 
contribution outside the matter, $P$ is a constant with 
temporal dimension, and $\kappa_l$ and 
$\kappa_l^{\prime}$ are defined as 
\begin{equation}
\kappa_l=\sum^{l-2}_{s=1}{1 \over s}+{2l^2+5l+4 \over l(l+1)(l+2)}, 
\label{kappa1}
\end{equation}
and 
\begin{equation}
\kappa_l^{\prime}=\sum^{l-1}_{s=1}{1 \over s}+{l-1 \over l(l+1)}. 
\label{kappa2}
\end{equation}
In black hole perturbation, the same tail corrections with 
multipole moments induced by linear perturbations have been obtained 
\cite{ps}. 

Equation ($\ref{waveform}$) does not apparently agree 
with Eqs.($\ref{blanmoment}$) and ($\ref{blanmoment2}$), 
because definitions of the moments 
$\{\tilde M_L, \tilde S_L\}$ and $\{M_L, S_L\}$ are different. 
We shall show below the equivalence between our expression and 
that of Blanchet \cite{blan95}, by counting the contribution 
from nonlinear terms like $M\times M_L$ or $M\times S_L$ 
in the source $\tilde\tau^{\mu\nu}$.

\subsection{Contributions from the nonlinear sources} 

In order to evaluate the tail of the waveform produced 
by the nonlinear sources in $\tilde\tau^{\mu\nu}$, it is enough 
to use the flat Green's function 
\begin{eqnarray}
G_{0}(x,x')&=&-i \sum_{lm} \int d\omega {\omega \over \pi}
\Bigl(e^{-i\omega t}h_l(\omega r)Y_{lm}(\Omega) 
e^{i\omega t'}j_l(\omega r')Y_{lm}^{*}(\Omega ')\theta(r-r') \nonumber\\
&&\hspace{3cm}+e^{-i\omega t}j_l(\omega r)Y_{lm}(\Omega) 
e^{i\omega t'}h_l(\omega r')Y_{lm}^{*}(\Omega ')\theta(r'-r) \Bigr) , 
\end{eqnarray}
where $j_l$ and $h_l$ are the first kind of the spherical Bessel 
function and the spherical Hankel function, respectively. 

We consider the asymptotic form of the following retarded integral 
\begin{eqnarray}
\Box^{-1}\Bigl[ {\hat n_L \over r^k} F(t-r) \Bigr]&&=
\int d^4x' G_0(x,x') \Bigl[ {\hat n'_L \over r'^k} F(t'-r') \Bigr]
\nonumber\\
&&\to -i \int d\omega \omega e^{-i\omega t} h_l(\omega r) F_{\omega} 
\hat n_L \int^r_0 r'^2 dr' j_l(\omega r') {e^{i \omega r'}\over r'^k} 
\nonumber\\
&&\hspace*{6cm}\mbox{for large r} , 
\label{formula0}
\end{eqnarray}
where the hat denotes the symmetric and traceless part of tensor
products and we defined $F_{\omega}$ as
\begin{equation}
F_{\omega}={1 \over 2\pi} \int dt \, e^{i\omega t} F(t) . 
\label{fomega}
\end{equation}
Thus, the contribution from nonlinear sources can be evaluated 
by using the formula 
\begin{eqnarray}
\Box^{-1} \Bigl[ {\hat n_L \over r^k}F(t-r) \Bigr] 
&=&-2^{l+1} \lim_{\lambda \to 0} \Bigl( \sum^{\infty}_{n=0} 
{(l+n+1)!\Gamma(-k+l+3+2n-\lambda) \over n! (2l+2n+2)!} \Bigr) 
\nonumber\\
&&\times{\hat n_L \over r}\mathop{F}^{(k-3)}\!\!(t-r) +O(r^{-2}) , 
\label{formula3}
\end{eqnarray}
where we used 
\begin{equation}
j_l(z)={\sqrt{\pi} \over 2} \sum^{\infty}_{n=0} {(-)^n (z/2)^{l+2n} 
\over n! \Gamma(l+n+3/2)} . 
\end{equation}

Since we consider both $M\times M_L$ and $M\times S_L$ in the 
nonlinear source, we have to treat the following type of 
retarded integral 
\begin{eqnarray}
\Box^{-1}\Bigl[ \partial_i({1 \over r}) \hat\partial_Q 
\Bigl({F(t-r) \over r} \Bigr)\Bigr]&=&(-)^{q+1} 
\sum^q_{j=0}{ (q+j)! \over 2^j j! (q-j)! } \nonumber\\ 
&&\times\Box^{-1}\Bigl[ \Bigl(\hat n_{iQ}+{q \over 2q+1}\delta_{i<a_q} 
\hat n_{Q-1>} \Bigr) {1 \over r^{j+3}} \mathop{F}^{(q-j)}\!\!(t-r) \Bigr] .
\nonumber\\
&&
\label{c201}
\end{eqnarray}
Using Eq.($\ref{formula3}$), we obtain 
\begin{eqnarray}
&&\Box^{-1} \Bigl[ \partial_i ({1 \over r}) \hat\partial_Q 
\Bigl( {F(t-r) \over r} \Bigr) \Bigr] \nonumber\\
&&={(-)^q \over 2(q+1)} \Bigl[ \hat n_{iQ}-{q+1 \over 2q+1}
\delta_{i<a_1} \hat n_{Q-1>} \Bigr] {1 \over r} 
\mathop{F}^{(q)}\!(t-r)+O(r^{-2}) , 
\label{c2}
\end{eqnarray}
where, in the limit $\lambda \to 0$ of Eq.($\ref{c2}$), poles 
of gamma functions cancel out in total and the finite values are obtained. 
This is same with the expression (C2) obtained by Blanchet \cite{blan95}. 

Applying Eqs.($\ref{c2}$) to the nonlinear source $\tilde\tau^{\mu\nu}$, 
we can evaluate its contribution to the waveform. 
Together with Eq.($\ref{waveform}$), we obtain the waveform 
in the same form as Eq.($\ref{tailwave}$).

\section{Discussion}

By considering the propagation of gravitational waves generated by 
slow motion sources in the Coulomb type potential, we have derived 
the formula ($\ref{waveform}$) for gravitational waves including tail. 
Our equation ($\ref{waveform}$) has been shown to be same as 
Eqs.($\ref{blanmoment}$)-($\ref{kappa2}$) in the previous work 
$\cite{blan95}$. 

We would like to emphasize two points on Eq.($\ref{waveform}$): 
First, in deriving Eq.($\ref{waveform}$), spherical Coulomb functions 
are used, since we use the wave operator ($\ref{box2}$) 
which take account of the Coulomb-type potential $M/r$. 
As a consequence, $\ln (v/2M)$ appears naturally in Eq.($\ref{tailwave}$). 
This is in contrast with Blanchet and Damour's method, 
where an arbitrary constant with temporal dimension $P$ appears 
in the form of $\ln (v/P)$. 
Our derivation shows that the main part of the tail, 
which needs the past history of the source only through $\ln{v}$, 
is produced by propagation in the Coulomb-type potential. 
The reason for saying the main part of the tail is as follows: 
Only the $log$ term has a hereditary property expressed as the integral 
over the past history of the source, since the constants $\kappa_l$ 
and $\kappa_l^{\prime}$ represent merely instantaneous parts 
after performing the integral under the assumption that 
the source approaches static as the past infinity. 
Our method allows us to study the effect on the waveform formula 
by the modification of the Coulomb type potential near 
the material source.  This effect might be important to make 
the comparison between theory and observation. We shall discuss this 
in future. 

The second point relates with physical application: 
At the starting point of our derivation, the Fourier representation 
in frequency space has been used. 
Such a representation seems to simplify the calculation of gravitational 
waveforms from compact binaries in the quasi-circular orbit, 
since such a system can be described by a characteristic frequency. 
Applications to physical systems will be also done 
in the future. 

\bigskip

\acknowledgements 
We would like to thank T. Nakamura for fruitful suggestion 
and discussion. 
One of us (H.A.) also would like to thank M. Sasaki, M. Shibata 
and T. Tanaka for useful discussion. 
This work is supported in part by Soryushi Shogakukai (H.A.) 
and in part by the Japanese Grant-in-Aid for Science Research fund 
of Ministry of Education, Science and Culture No.09640332 (T.F.).

\end{document}